\documentclass[apjl]{emulateapj}
\usepackage[]{natbib}
\usepackage{color}
\definecolor{amber}{rgb}{1.0, 0.75, 0.0}

\usepackage [english]{babel}
\usepackage [autostyle, english = american]{csquotes}
\MakeOuterQuote{"}

\usepackage[breaklinks,colorlinks, urlcolor=blue,citecolor=blue,linkcolor=blue]{hyperref}

\shorttitle{Clump Clusters and Giant Clumps in High Redshift Galaxies}
\shortauthors{Behrendt et al.}

\begin{document}

\title{Clusters of Small Clumps Can Explain The Peculiar Properties \\ of Giant Clumps in High-Redshift Galaxies}


\author{M. Behrendt\altaffilmark{1,2}, A. Burkert\altaffilmark{1,2,3} and M. Schartmann\altaffilmark{4}}

\email{E-mail: mabe@mpe.mpg.de}
\altaffiltext{1}{Max Planck Institute for extraterrestrial Physics,\\ PO box 1312, Giessenbachstra\ss e, D-85741 Garching, Germany }
\altaffiltext{2}{University Observatory Munich, \\ Scheinerstra\ss e 1, D-81679 Munich, Germany}
\altaffiltext{3}{Max Planck Fellow}
\altaffiltext{4}{Centre for Astrophysics and Supercomputing,\\ Swinburne University of Technology, Hawthorn, Victoria 3122, Australia}

\submitted{Accepted version February 5, 2016}

\begin{abstract}
Giant clumps are a characteristic feature of observed high-redshift disk galaxies. We propose that these kpc-sized clumps have a complex substructure and are the result of many smaller clumps self-organizing themselves into clump clusters (CC). This bottom-up scenario is in contrast to the common top-down view that these giant clumps form first and then sub fragment. Using a high resolution hydrodynamical simulation of an isolated, fragmented massive gas disk and mimicking the observations from \citet{2011ApJ...733..101G} at $z \sim 2$, we find remarkable agreement in many details. The CCs appear as single entities of sizes $R_{\mathrm{HWHM}} \simeq 0.9-1.4$ kpc and masses $\sim 1.5-3 \times 10^9 \ \mathrm{M_{\sun}}$ representative of high-z observations. They are organized in a ring around the center of the galaxy. The origin of the observed clumps' high intrinsic velocity dispersion $\sigma_{\mathrm{intrinsic}} \simeq 50 - 100 \ \mathrm{km \ s^{-1}}$ is fully explained by the internal irregular motions of their substructure in our simulation. No additional energy input, e.g. via stellar feedback, is necessary. Furthermore, in agreement with observations, we find a small velocity gradient $V_{\mathrm{grad}} \simeq 8 - 27 \ \mathrm{km \ s^{-1} \ kpc^{-1}}$ along the CCs in the beam smeared velocity residual maps which corresponds to net prograde and retrograde rotation with respect to the rotation of the galactic disk. The CC scenario could have strong implications for the internal evolution, lifetimes and the migration timescales of the observed giant clumps, bulge growth and AGN activity, stellar feedback and the chemical enrichment history of galactic disks. 
\end{abstract}





\section{Introduction}
Typical characteristics of observed high redshift ($z \sim 1-3$) star-forming galaxies are their large baryonic cold gas fractions \citep{2008ApJ...673L..21D, 2010ApJ...713..686D, 2008ApJ...680..246T, 2010Natur.463..781T, 2013ApJ...768...74T} and high velocity dispersion, their irregular morphology and a few kpc-sized clumps containing baryonic masses of $\gtrsim 10^8-10^9 \ \mathrm{M_{\odot}}$ \citep{2004ApJ...604L..21E, 2005ApJ...634..101E,2007ApJ...658..763E, 2009ApJ...706.1364F,2011ApJ...731...65F, 2011ApJ...739...45F, 2011ApJ...733..101G, 2014ApJ...785...75G}. The common understanding is that these higher gas fractions and densities lead to gravitationally unstable disks with a Toomre parameter $Q < 0.7$ \citep{1964ApJ...139.1217T, 1965MNRAS.130...97G, 2015MNRAS.448.1007B} that fragment into a few kpc-sized objects \citep[][and references therein]{Bournaud:2016je}. This picture is supported by the detection of massive clumps in observations and from cosmological simulations. Linear perturbation theory indeed predicts a dominant growing wavelength of order of several 100 pc to kpc \citep[e.g.][]{Dekel:2009bn, 2011ApJ...733..101G}. \citet[][]{2015MNRAS.448.1007B} however showed recently that this wavelength determines the initial sizes of a few axisymmetric rings growing from inside-out instead of a few kpc-sized clumps if initially the densities in the mid-plane are sufficiently resolved (see also \citealt{2014ApJ...780...57B} and Figure 6 in \citealt{Bournaud:2016je}). These rings break up into many clumps after they collapsed onto pc-scales. The typical clump ensemble in a simulation of a massive disk with $\sim 3 \times 10^{10} \ \mathrm{M_{\sun}}$ is initially dominated by clumps with average masses of $\sim 2 \times 10^7 \ \mathrm{M_{\odot}}$ and a typical radius $R \sim 35$ pc and later on most of the mass resides in a population of clumps with $\sim 2 \times 10^8 \ \mathrm{M_{\odot}}$ and a radius of $R \sim 60$ pc  (M. Behrendt et al., 2016, in preparation). Interestingly, a similar mass range has been found in the studies of \citet{2015MNRAS.453.2490T} where the clumps fragment from spiral features produced at $Q \sim Q_{\mathrm{crit}}$. Later on, the clumps in our simulation quickly form groups on several 100 pc to kpc-scales which we will call clump clusters \footnotemark[1] (CC).
	\footnotetext[1]{The term "clump clusters" was used in the literature previously to describe a conglomeration of kpc-sized structures seen at high redshifts. It was later on abandoned and replaced by the notation "clumpy galaxies" \citep[see the explanation in][]{2005ApJ...627..632E}.}\\
In this letter we explore the question whether these CCs could resemble the properties of the giant clumps observed in redshift 2 galaxies. An important role plays the instrumental resolution which spatially smears-out the information on kpc-scales. Hints for substructure are however mainly given by local gas-rich disk galaxies which can be observed with much higher resolution. The DYNAMO survey identified local galaxies with very similar properties as found at $z \sim 2$  \citep{2010Natur.467..684G, 2014ApJ...790L..30F, 2014MNRAS.442.3206B} containing clumps with typical diameters $< d_{\mathrm{clump}} > \sim 0.6$ kpc \citep{2015IAUGA..2256258F}. \\
We use a higher resolution version of the simulation presented in  \citet[][]{2015MNRAS.448.1007B} and compare clump cluster properties with the giant clumps of the five luminous star-forming disk galaxies at $z \sim 2$ from \citet{2011ApJ...733..101G}. This pioneering work  for the first time presented detailed line profiles of individual giant clumps. The following list summarizes the observationally motivated questions that can be explained by our CC scenario.

\begin{enumerate}
  \item Do the giant clumps have substructure? Observationally, only one example of a very bright clump is found where a substructure is hinted in the velocity channel maps given the large beam smearing.
  \item What is the origin of the high intrinsic velocity dispersion $\sigma_{\mathrm{intrinsic}} \simeq 50 - 100 \ \mathrm{km \ s^{-1}}$ of the clumps? Often this is attributed to stellar feedback, but in the analysis of the observations no significant correlation between local velocity dispersion and star formation rates could be found.
  \item Are the giant clumps rotationally supported? Small velocity gradients are found along the clumps $V_{\mathrm{grad}} \simeq 10-30 \ \mathrm{km \ s^{-1} \ kpc^{-1}}$ in the velocity residual maps which corresponds to net prograde and retrograde rotation with respect to the rotation of the galaxy. When considering the beam-smearing effects on the kinematics \citet{2011ApJ...733..101G} concluded that these clumps are either pressure supported by high velocity dispersion \citep[see also][]{2013MNRAS.432..455D} or they are still undergoing collapse because of the small velocity gradients.
\end{enumerate}
In Section \ref{sec:Simulation} we describe the simulation code and the disk setup and how we mimick the observations. Section \ref{sec:Results} gives an overview of the disk evolution and addresses the single issues listed above. Finally, in Section \ref{sec:Summary and Conclusion} we summarize the results and derive implications.

\section{Simulation}
\label{sec:Simulation}

\subsection{Code and disk setup}
We run the simulation with the hydrodynamical AMR code RAMSES \citep{2002A&A...385..337T} in a box of 48 kpc and with a maximum resolution $\Delta x = 2.9$ pc. The hydrodynamical equations describing the evolution of the self-gravitating gas disk with an isothermal equation of state are solved by the HLL Riemann solver and the MinMod slope limiter \citep{2006A&A...457..371F}. A static dark matter halo is added as an external potential. To sufficiently resolve the mid-plane gas density distribution we represent the Jeans length by at least $N_{\mathrm{J}} = 19$ grid cells at every resolution level and the refinement criterium stops at maximum resolution. To avoid artificial fragmentation also for higher densities at 2.9 pc resolution we add a pressure floor in order to assure  $N_{\mathrm{J}} =7$ grid elements per Jeans length which leads to a lower limit for the clump radius of 10.3 pc  \citep{1997ApJ...489L.179T, Agertz:2009wd, Bournad:2010et}.
The same disk model as in \citet{2015MNRAS.448.1007B} is used. We adopt an exponential gas disk with total mass $M_{\mathrm{disk}} = 2.7 \times 10^{10} \ \mathrm{M_{\mathrm{\sun}}}$, scale length $h = 5.26 \ \mathrm{kpc}$ and outer radius $R_{\mathrm{disk}} = 16 \ \mathrm{kpc}$. The Toomre parameter is $Q<0.7$ within $10.5 \ \mathrm{kpc}$ and the disk is therefore unstable to axisymmetric modes in this radius regime. The dark matter halo has the density profile of \citet{Burkert:1995jr} with $M_{\mathrm{DM}} = 1.03 \times 10^{11} \ \mathrm{M_{\sun}}$ within 16 kpc. The isothermal temperature of $10^4$ K represents the typical micro-turbulent pressure floor and keeps the initial vertical density distribution stable until the disk fragments.

\subsection{Mimicking the observations}
We compare the results with the clump properties of the H$\alpha$ observations of five disk galaxies from \citet{2011ApJ...733..101G} at $z \sim 2.2-2.4$. The instrumental angular resolutions of $0\arcsec .18 - 0\arcsec .25$ correspond to $\mathrm{FWHM} \simeq 1.5-2.1 \ \mathrm{kpc}$. We "observe" the simulated galaxy at an inclination $i=60^\circ$ which is the most likely value for a random orientation and construct line-of-sight (LOS) maps of the spatial and kinematic components. The resolution of the observational instrument is mimicked by convolving the LOS surface density with a 2D Gaussian of FWHM = 1.6 kpc (Section 3.2). The result can be interpreted as a map of the molecular gas surface density since the majority of the gas mass ($\sim 76 \%$) resides in clumps with surface densities $\gg 100 \ \mathrm{M_{\odot} \ pc^{-2}}$. This corresponds to the observed H$\alpha$ maps due to the usually adopted linear "Kennicutt-Schmidt" relation from \citet{2013ApJ...768...74T} (PHIBBS calibration), see also \citet{2014ApJ...785...75G}. For the kinematic analysis we take the mass-weighted LOS velocity information of the simulation. \citet{2013ApJ...768...74T} found that the ratio of rotational velocity to local velocity dispersion in CO agree to first order with ratios obtained from  H$\alpha$ of similar star forming galaxies. The LOS velocities of the clump regions in Section (3.3) are binned into "channels" of width 34 $\mathrm{km \ s^{-1}}$ and spatially convolved. To obtain the intrinsic clump velocities the beam-smeared LOS velocity of a rotating featureless exponential disk model is subtracted. We do not convolve with the instrumental spectral resolution of $\mathrm{FWHM} = 85 \ \mathrm{km \ s^{-1}}$ since this contribution is already removed in the observational values we compare with.

\begin{figure*}
\centering
\includegraphics{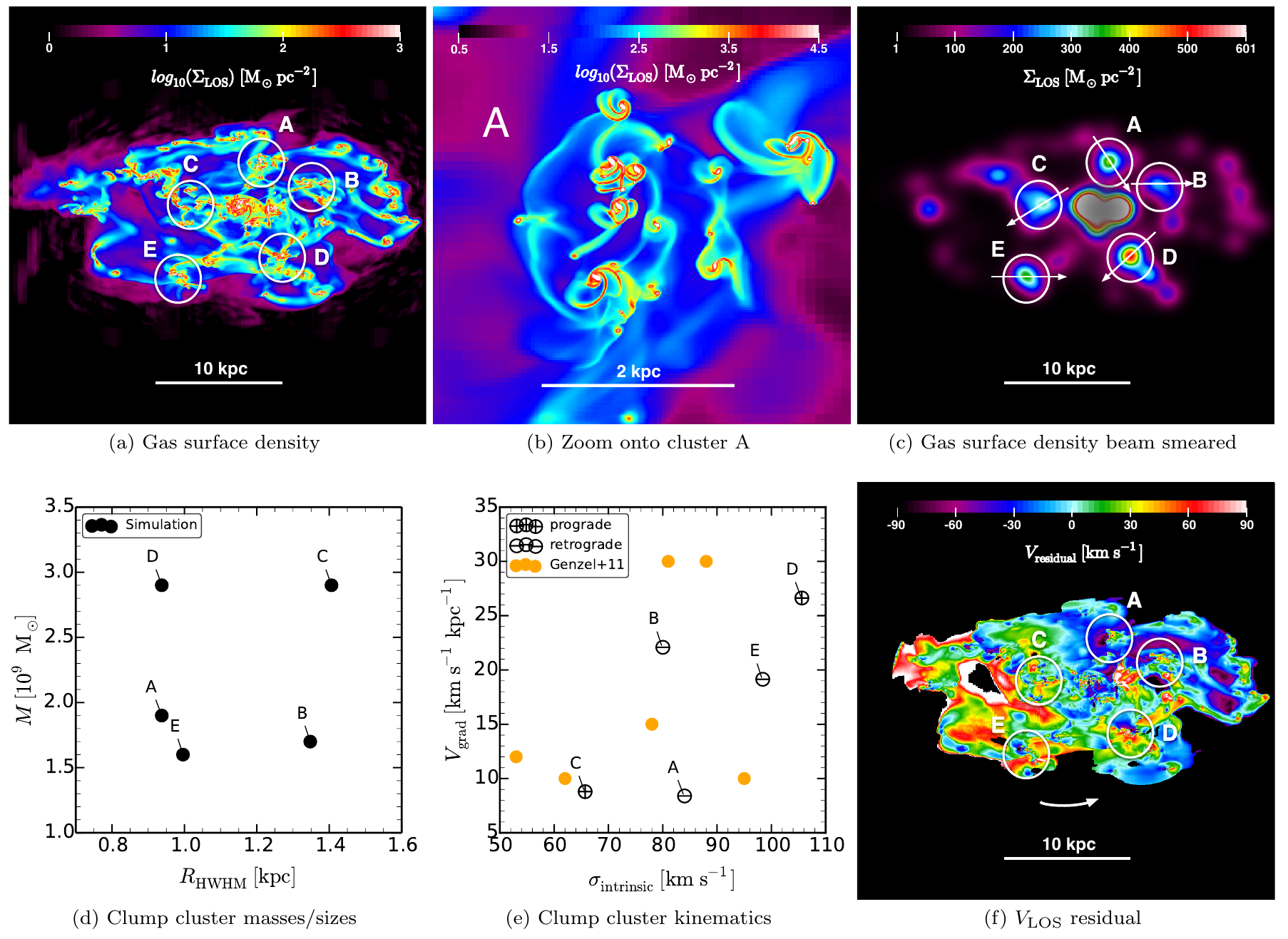}

\caption{(a) The LOS surface density of the galaxy at an inclination $i=60^{\circ}$ (limited to a maximum of $10^3 \ \mathrm{M_{\odot} \ pc^{-2}}$). The analyzed clusters are marked with circles and labeled (A, B, C, D, E). (b) The zoom-in face-on surface density map of cluster A limited to the maximum surface density $10^{4.5} \ \mathrm{M_{\odot} \ pc^{-2}}$. (c) The beam smeared LOS surface density ($\mathrm{FWHM} = 1.6$ kpc) of the galaxy at $i=60^{\circ}$. The arrows represent the measured cuts of the velocity gradients and point into the direction from blue to redshifted. $V_{\mathrm{grad}}$ is calculated within $R_{\mathrm{HWHM}}$ in Equation (\ref{eq:vgrad}). The center is intentionally shaded for better visualisation of the clump structure in the disk region. (d) The masses of the simulated CCs and their radii $R_{\mathrm{HWHM}}$. (e) The velocity gradients along the clusters (arrows in Figure 1c) within $R_{\mathrm{HWHM}}$ and their intrinsic velocity dispersion are compared with the observations \citep{2011ApJ...733..101G}. Their rotational direction is given by the plus symbol for prograde and the minus symbol for retrograde spin with respect to the rotation of the galaxy. (f) LOS velocity residual maps (limited velocity range) of the galaxy at $i=60^{\circ}$ for surface densities $\Sigma_{\mathrm{LOS}} > 4 \ \mathrm{M_{\odot} \ pc^{-2}}$ and within the disk radius of 16 kpc.   \label{fig:maps}}
\end{figure*}

\section{Results}
\label{sec:Results}

\subsection{Overview of the disk evolution}
The simulation evolves like in \citet{2015MNRAS.448.1007B}. The main difference in the new simulation is its higher spatial resolution which in turn increases the resolution of the CCs that are the focus of this letter. In addition, we were able to reduce the pressure floor to a more realistical value. The unstable disk fragments into rings from inside-out in excellent agreement with the local fastest growing wavelength. The rings subsequently break up into hundreds of clumps, identified with a clump finder \citep{2014MNRAS.445.4015B} and considering gas densities $\mathrm{n_{H} \geq 100 \ cm^{-3}}$. The clump statistic will be discussed in great details in a forthcoming paper (M. Behrendt et al., 2016, in preparation). The clumps initially have typical masses of several $ \times 10^7 \ \mathrm{M_{\sun}}$ and radii around 35 pc. They later on evolve by merging with other clumps and re-organize themselves into large clump clusters. The fragmented disk (at 655 Myr) can be seen in the LOS surface density map in Figure 1a. In this simulation we do not include stellar feedback. We however refer to the study and discussion in \citet{2014ApJ...780...57B} who have shown that mainly the clumps below a mass of a few $10^7 \ \mathrm{M_{\sun}}$ are short-lived and effected by stellar feedback processes while the more massive clumps can survive several hundreds of Myr which is long enough for clump clusters to form, as discussed in the next section. In future work we will include stellar feedback in order to investigate its effect on the evolution.

\subsection{Clump clusters appear as kpc-sized clumps}
The clumps organize themselves into clusters on several 100 pc to kpc scales. These clump clusters represent groups of individual clumps with diameters $\sim$ 100 pc. As an example, a zoom-in onto the Cluster A in the evolved disk (655 Myr) is shown in Figure 1b. The region has a radius of $\sim 1.25 \ \mathrm{kpc}$ in the face-on view. When we convolve the LOS surface density map with a Gaussian filter of FWHM = 1.6 kpc  the substructure of the CCs is completely smeared out (Figure 1c). They now appear as single entities, arranged into a ring with a radius of 4 - 7.7 kpc around the center of the galaxy. At t=655 Myr we also find that clump clusters merge in the center of the galaxy, leading to the formation of a bulge component. The details of bulge formation by CCs will be investigated in a forthcoming paper (M. Behrendt et al., 2016, in preparation). Here, we focus on the prominent CCs, labeled A,B,C,D,E in the galactic disk region. Since the disk is fragmenting from inside-out these clusters were build from clumps that formed after 300 Myr. The beam smeared CC's surface densities peak at $\Sigma_{\mathrm{LOS}} = 300-600 \ \mathrm{M_{\odot} \ pc^{-2}}$ and the more elongated clusters (B, C) have somewhat smaller surface densities in the convolved map compared to the clusters with more concentrated substructure (A, D, E). Their large radii of $R_{\mathrm{HWHM}} = 0.95 - 1.4$ kpc (Figure 1d) are similar to the identified clumps in \citet{2011ApJ...733..101G}. The CC's total masses are $1.6-3 \times \ 10^9 \ \mathrm{M_{\sun}}$. The typical clumps in \citet{2011ApJ...733..101G} have masses a few times $10^9 \ \mathrm{M_{\odot}}$ and the most extreme clumps masses of  $\sim 10^{10} \ \mathrm{M_{\odot}}$. The difference to our result can by explained by the around three times larger total baryonic mass in their observed galaxies. 

\subsection{Origin of the intrinsic high velocity dispersion}
The measured intrinsic velocity dispersion of the five CCs is shown in Figure 1e. Before we construct the integrated spectrum of the cluster regions we take the LOS velocity and bin it into "channel" maps of width 34 $\mathrm{km \ s^{-1}}$ which we spatially convolve with a Gaussian of FWHM = 1.6 kpc. Then we subtract the beam-smeared rotation of the smoothed galactic disk model to extract the velocity imprint of the clumps. The remaining integrated residual velocities of a cluster region are normalized to the maximum "intensity" (LOS surface density). As shown in Figure 1e our CCs have velocity dispersions of $\sigma_{\mathrm{intrinsic}} \simeq 65 - 105 \ \mathrm{km \ s^{-1}}$ in remarkable agreement with the observations. These high values are a result of the internal irregular motions of their substructures. The original (unsmeared) data gives almost the same velocity dispersion with only $1-5 \ \mathrm{km \ s^{-1}}$ difference. 

\subsection{Spin of the kpc-sized clumps}
In not enough resolved numerical simulations of gas-rich disks the kpc-sized clumps are fast rotating  and supported by internal centrifugal forces \citep[e.g.][]{2004ApJ...611...20I, 2004A&A...413..547I, Dekel:2009bn, 2010ApJ...719.1230A}. This is in contradiction to the observations that indicate dispersion-dominated clumps, being stabilized by random motion rather than rotation \citep[see also][]{2013MNRAS.432..455D}. The high resolution simulations presented in \citet[][]{2012MNRAS.420.3490C} indeed show a rich spectrum of substructures  \citep[see also][]{Bournaud:2016je}, similar to our simulation. They argue that the internal supersonic turbulence dominates the kinematics of the giant clumps and induces the break-up into sub-units. We instead conclude that small clumps form first, organize themselves to giant clusters which build the substructure that regulates the cluster dynamics. Following Genzel et al. (2011) we measure velocity gradients along cuts of largest gradients 
\begin{equation}
\label{eq:vgrad}
 V_{\mathrm{grad}} = \frac{(v_{\mathrm{max}} - v_{\mathrm{min}})_{\mathrm{residual}}}{2 \sin(i) R_{\mathrm{HWHM}}} 
\end{equation}
of the CCs of our beam smeared residual maps (inclination corrected) which we indicate with arrows in Figure 1c. The gradients range in between $V_{\mathrm{grad}} \simeq 8-27 \ \mathrm{km \ s^{-1} \ kpc^{-1}}$ (Figure 1e) which is again in surprisingly good agreement with the observations. The self-rotation of the individual clumps within a CC is completely washed out and only a gradient over the whole cluster region remains which has several reasons. For better understanding we show the un-smeared LOS residual velocities in Figure 1f where the clusters can be identified as "Rotating Islands" (A, B, D, E). They appear as large blue and red shifted areas with small substructures of high velocities representing the individual, spinning clumps. Cluster C shows only a modest imprint of a velocity gradient with $\sim 8 \ \mathrm{km \ s^{-1} \ kpc^{-1}}$ since its substructures are "coincidently" close together. The other clusters  rotate either slowly or faster around their centers of mass and are continuously perturbed by encounters with other CCs. We could not find any correlation between the kinematics and their masses or radii. Clusters with a larger $V_{\mathrm{grad}}$ have a tendency to also have an increased $\sigma_{\mathrm{intrinsic}}$ in our snapshot, however, the statistics is too low to argue that it is significant. The observed clumps also show prograde and retrograde velocity gradients with respect to the rotation of the galaxy. A giant clump which formed due to gravitational instability in a sheared disk should however have an angular momentum vector pointing into the same direction as its host galaxy. This coordinated rotation is indeed seen in  the clumps that formed initially by gravitational disk instabilities. The situation is however different for our CCs as they continuously interact or merge with clumps. The clusters A,B,E rotate retrograde while C and D rotate prograde (indicated by the direction of the arrow from blue to redshifted velocities in Figure 1c and by the symbols in Figure 1e). 

\section{Summary and Conclusions} 
\label{sec:Summary and Conclusion}
We compared a high-resolution simulation of a clumpy massive gas disk with the $z \sim 2$ galaxies of \citet{2011ApJ...733..101G}. The Toomre unstable $Q < Q_{\mathrm{crit}}$ disk naturally evolves into a large number of clumps, initially with an average mass $2 \times 10^7 \ \mathrm{M_{\sun}}$  ($R=35$ pc) and later on most of the mass in mergers of $\sim 2 \times 10^8 \ \mathrm{M_{\odot}}$ with a radius of $R \sim 60$ pc (M. Behrendt et al., 2016, in preparation). They subsequently self-organize into several 100 pc to kpc-sized clump clusters (CCs) in a ring-like distribution. We analyze the fully fragmented disk at an inclination of $i=60^{\circ}$. By mimicking the observations we find the following results:
 
\begin{enumerate}
  \item In the beam-smeared disk (FWHM = 1.6 kpc) the small-scale substructure disappears and only a few giant clumps are visible with $R_{\mathrm{HWHM}} \simeq 0.9-1.4$ kpc and  masses of $\sim 1.5-3 \times 10^9 \ \mathrm{M_{\sun}}$. They are organized in a ring with a radius of 4 - 7.7 kpc around the center of the galaxy. The giant clumps are actually clump clusters and show a rich substructure on pc scales. The model galaxy has around three times less baryonic mass than the observed galaxies in \citet{2011ApJ...733..101G} which is reflected in 3 times less massive clump clusters.
  \item The high intrinsic velocity dispersion $\sigma_{\mathrm{intrinsic}} \simeq 65-105 \ \mathrm{km \ s^{-1}}$ of the CCs is caused by their subclump's high  irregular motions. This is in contrast to previous assumptions that attributed the dispersion to turbulence, generated by stellar feedback. We tested the effect of beam smearing on the inferred velocity dispersions and find no significant differences, indicating that these signatures can be used as realistic indicators of the CC kinematics. The observed high dispersion of massive clumps in gas-rich galaxies might therefore be indirect evidence for a cluster of weakly bound substructures and a characteristic property of CCs. This is also in agreement with the finding of \citet{2011ApJ...733..101G} that there does not exist a correlation between their dispersion of $\sigma_{\mathrm{intrinsic}} \simeq 53 - 95 \ \mathrm{km \ s^{-1}}$ and the star formation rates.   
  \item The clump clusters show small velocity gradients $V_{\mathrm{grad}} \simeq 8-27 \ \mathrm{km \ s^{-1} \ kpc^{-1}}$ which corresponds to net prograde or retrograde rotation with respect to the galaxy. The larger values correspond to faster rotating CCs and the smaller either to slowly spinning clusters or the substructure is "coincidently" close together to appear as a giant clump when beam-smeared.
 \end{enumerate}
We demonstrated that clump clusters can explain many observed properties of giant clumps at high-redshift. If the observed unresolved massive clumps indeed are ensembles of dense subclumps, this has strong implications for any model that infers their evolution. Kpc sized clumps are expected to migrate to the disk center via dynamical friction and tidal torques on a few orbital timescales where they contribute to the formation of a bulge \citep{1999ApJ...514...77N, 2004ApJ...611...20I, 2004A&A...413..547I, 2006ApJ...645.1062F, 2006Natur.442..786G, 2008ApJ...687...59G, 2011ApJ...733..101G, 2008ApJ...688...67E, 2007ApJ...658..960C, 2009ApJ...707L...1B, Dekel:2009bn, Ceverino:2010eh, Bournaud:2016je}. The CC scenario could have a strong effect on the estimated migration timescale of dense gas into centers of gas-rich galaxies which can have strong influence on the feeding of central black holes and AGN activity. Star formation and stellar feedback processes should also be strongly affected by the substructure \citep{2013MNRAS.432..455D} and their chemical enrichment history. Here we focused on the structure of CCs and their observational properties. However, CCs also have an interesting and complex evolution. They for example are exchanging their substructure or even disperse and reform. This will be discussed in detail in a subsequent paper.

\acknowledgments
We thank the referee for constructive comments that improved the quality of the manuscript. We are grateful to Philipp Lang, Lucio Mayer, Go Ogiya, Michael Opitsch and Valentina Tamburello for fruitful discussions. We also thank Fr\'ed\'eric Bournaud and Avishai Dekel for valuable comments. The computer simulations were performed on the HPC system HYDRA at the Rechenzentrum Garching (RZG) of the Max Planck Gesellschaft.

\end{document}